 \documentstyle[11pt,fleqn]{article}
 \newcommand{\eqnsection}{
 \renewcommand{\theequation}{\thesection.\arabic{equation}}
 \makeatletter
 \csname $addtoreset\endcsname
 \makeatother}
\begin{document}
 \eqnsection
 \title{Universality of the universal $R$-matrix and
  applications to quantum integrable systems }
 \author{
 Anjan Kundu  \\
 Fachbereich 17--Mathematik/ Informatik, \
 GH--Universit"at Kassel \\
 Holl"andische Str. 36 \, 34109 Kassel, Germany
 \\{\it Permanent address}:\\
  Saha Institute of Nuclear Physics, AF/1 Bidhan Nagar,
  \\ Calcutta 700 064,India.
 }
 \date{}
 \maketitle
 \begin{abstract}
 %------------------------------------------------------------
 Results obtained by us are overviewed from a general set up.
  The universal $R$-matrix
 is exploited   to
  obtain various important relations and structures involved in
  quantum group algebra, which are
  used  subsequently for generating
 different classes of quantum integrable systems through a systematic scheme.
 This recovers known models as well as discovers a series of new ones.

 \end{abstract}
 %%%%%%%%%%%%%%%%%%%%%%%%%%%%%%%%%%%%%%%%%%%%%%%%%%%%%%%%%%%%%%%%%%
 %\setcounter{equation}{0}
 \noindent  {\bf Introduction}:

    Our basic aim is to construct a systematic scheme for generating
    different classes of quantum integrable systems (QIS) starting from some
  well known and universal structure related to quantised  algebra (QA).
 We also
   attempt to answer the following intriguing
 questions  related to QIS like,
 \\ $\circ$ \ How a single   QA may be related to
  diverse forms of nonlinar integrable models.
 \\ $\circ$ \ What is the criterion of  nonlinearity that
  makes a  system
  integrable.\\ $\circ$ \
  What is the `commonness' between seemingly widely diffrent  models like
   sine-Gordon (SG), derivative nonlinear Schrodinger (DNLS) ,
  massive Thirring model (MTM) etc. (all of them share the same quantum
  $R$-matrix !).

 Our starting point in achieving our goal would be an universal object -
 the universal ${ \cal R}$-matix $ \in U_q(g) \otimes U_q(g)$,
  which satisfies the 'universal' Yang-Baxter equation (UYBE):.
 $ \ \   R_{{\hat 1\hat 2}}~R_{{\hat 1\hat 3}} ~R_{{\hat 2\hat3}}
 ~=~~R_{{\hat 2\hat 3}}
     ~R_{ {\hat 1 \hat3}} ~R_{{ \hat 1 \hat 2}} ~,
 \ \ $  where $\hat i$ indicates the infinite dimensional operator (quantum)
 space.
 We show first that the universal $ {\cal R}$-matrix is universal
 enough to yield the  Faddeev-Reshitikhin-Takhtajan (FRT) algebra [1]
 itself along with the explicit forms of  $ R^{\pm}, L^{\pm}$
 involved in this algebra. Since our construction relates further
 these objects  to  quantum integrable
 systems through Yang-Baxterisation,
   forms of  universal ${\cal R}$-matrix in a sense
 answer ultimately for the generated   classes of integrable
 systems. Moreover  it is demonstrated that
  the $ {\cal R}$-matrix
 and the UYBE   give also
 the explicit  coproduct structures
 as well as the intertwinning relation between  coproducts.
 Frequently, all these important objects are defined seperately and introduced
 as independent entries, though in many occations  their relation with the
 ${\cal R}$-matrix
  has been mentioned.  We intend to derive here  these relations in a
 consistent way and use them for our subsequent application to integrable
 systems.

 Not to make  things abstract let us keep before us
  the universal ${\cal R}$-matrix  related to $U_q(sl(2)) $
   ,which has relatively
  simpler form:
 \begin{equation}
 {\cal     ~R}=~q^{2(H \otimes H)} \  Exp_q (q^H X^+ \otimes q^{-H} X^-)
 \end{equation}
 %(1)
  Note now that all finite-dimensional representations of these matrices
  are upper-triangular  in form. Therefore for concreteness  we denote
  such universal $ {\cal R}$-matrices
   as $ {\cal R}^+$ and the related equation
 as UYBE $(+++)$ :\
 $ i)\qquad
  R^+_{{\hat 1 \hat2}}~R^+_{{ \hat 1 \hat3}} ~R^+_{{ \hat 2 \hat3}}
  ~~=~~R^+_{{ \hat 2 \hat3}}
     ~R^+_{ { \hat 1  \hat 3}} ~R^+_{{  \hat 1  \hat 2}} ~. \
 $
 It is not difficult to see that if we define ${\cal R}^- = {\cal  P}
  ({\cal R}^+)^{-1}{\cal P}$,\ it would satisfy now the UYBE $(---)$: \
 $ii)\qquad
  R^-_{{\hat 1  \hat2}}~R^-_{{  \hat 1  \hat3}} ~R^-_{{  \hat 2  \hat3}}
  ~~=~~R^-_{{  \hat 2  \hat3}}
     ~R^-_{ {  \hat 1 \hat3}} ~R^-_{{ \hat 1 \hat2}} ~,
 $ \
 and will be lower-triangular  in form. More interestingly along with the
 above two universal equations another four
  UYBE of $ (+- -), (- ++), (++ -)$ and $(- -+)$ type would also  equally
 hold:
 $$iii)\qquad
  R^+_{{\hat 1 \hat2}}~R^-_{{ \hat 1 \hat3}} ~R^-_{{ \hat 2 \hat3}}
  ~~=~~R^-_{{ \hat 2 \hat3}}
     ~R^-_{ { \hat 1 \hat3}} ~R^+_{{ \hat 1 \hat2}} ~,
 \ \ \qquad
 iv) \qquad R^-_{{ \hat 1 \hat2}}~R^+_{{ \hat 1 \hat3}} ~R^+_{{ \hat 2 \hat3}}
  ~~=~~R^+_{{ \hat 2 \hat3}}
     ~R^+_{ { \hat 1 \hat3}} ~R^-_{{ \hat 1 \hat2}} ~,$$
 $$ v) \qquad
  R^+_{{ \hat 1
   \hat2}}~R^+_{{ \hat 1 \hat3}} ~R^-_{{ \hat 2 \hat3}} ~~=~~R^-_{{ \hat 2
    \hat3}}
     ~R^+_{ {
    \hat 1
    \hat3}} ~R^+_{{\hat 1\hat2}} ~,
                     \ \ \qquad vi) \qquad
  R^-_{{\hat 1 \hat2}}~R^-_{{ \hat 1 \hat3}} ~R^+_{{ \hat 2 \hat3}}
  ~~=~~R^+_{{ \hat 2 \hat3}}
     ~R^-_{ { \hat 1 \hat3}} ~R^-_{{ \hat 1 \hat2}} ~,
 $$
 which can be proved
  easily using the permutation property  of $ {\cal P}$.
   At the same time one also verifies easily
    the nonvalidity of $(+-+)$ type UYBE
   (holds only for triangular
  Hopf algebras: $ R_{{\hat 1 \hat2}}^{-1}= R_{{ \hat 2 \hat1}}$ ).
                 \\ \\
 \noindent  {\bf Reductions of UYBE and consequences}:

 Now we show that  different reductions  of the above UYBE's
 can yield different important relations
 relevant to quantum group.
 Here  by without {\it hat} numbers we denote
  $ 2\times2$ dimensional vector spaces reduced from
  infinite
 dimensional operator (quantum) spaces.
  Notice
  that the reduction $ {\hat 1}\rightarrow 1, {\hat 2}\rightarrow 2 , {\hat
   3}\rightarrow
    {\hat 3}$   yield interestingly the famous
   FRT algebra
  simply  as a consequence of the above UYBE's .
   Under this reduction the $(+++)$ UYBE reduces to
 \begin{equation}
  R^+_{{ 12}}~R^+_{1{\hat 3}} ~R^+_{2{\hat 3}} ~~=~~R^+_{2{\hat 3}}
     ~R^+_{1{\hat 3}} ~R^+_{{ 12}} ~,
 \end{equation}
 %(2)
 where $   R^+_{{ 12}}=  R^+_{BGR} $ is
  now a $4\times4$-matrix (braid group representation(BGR))
   and $ R^+_{1{\hat 3}}=
  L^+_{ 1({\hat 3})}$ and $ R^+_{2{\hat 3}}=L^+_{ 2({\hat 3})} $
    are $2\times2$ upper triangular matrices
  with operator
   valued elements. With these notations  eqn. (2)  clearly  turns into
  one of the ($(+++)$ type) FRT relations. The other
 FRT relations are similarly
  obtained  from UYBE  $(+- -),(++ -)$ etc.,
   yielding thereby the corresponding
  quantised algebra as the quadratic relations among the elements of
  $L^{\pm}$ acting in the quantum space ${\hat 3}$.
  Note that why an FRT relation of $(+ - +)$
  type does not hold
   is also answered naturally by the nonvalidity of such
  UYBE ,as mentioned above.

 On the other hand for the complementary reduction like
              $ {\hat 1}\rightarrow {\hat 1}, {\hat 2}\rightarrow {\hat 2}
               , {\hat 3}\rightarrow
    { 3}$  we may derive  the intertwining property of the universal
   $ {\cal R}$-matrix  as well as the explicit coproduct structures
  again as a consequence of the above UYBE's. For example UYBE $(+++)$ yields
 $
  R^+_{{\hat 1 \hat2}}~(L^{({ \hat 1})}_3 \cdot L^{({ \hat 2})}_3) =
   (L^{({ \hat 2})}_3 \cdot L^{({ \hat 1)}}_3)
  ~R^+_{{ \hat 1 \hat2}}
 $
 with the notation  $ R^+_{{ \hat 1}3} = L^{({ \hat 1})}_3 $ etc. ,where $
   L^{({ \hat 1})}_3 $ is now a lower triangular matrix with operator elements
  $ \{ l^{({ \hat 1})}_{ij} \}$ belonging to the quantum space ${\hat 1}$.
 It is not difficult to see that the above equation gives in a matrix
 form the intertwinning property: $  R^+_{{\hat 1 \hat2}}~ \Delta  (
  (\{ l_{ij} \}     )   =\tilde\Delta (          (\{ l_{ij} \} )
          R^+_{{\hat 1 \hat2}}~
  $ for all the elements $ \{ l_{ij} \}$ .  The explicit form of $ {\cal R}$>
also yields naturally the explicit expressions for
   coproducts;   relations
  for other elements being similarly obtained from equation of $(+- -)$ type.
 We may observe here that there is another interesting
  way to get the coproduct structure by using an important property of
   (2) i.e. ,
   if $ L^+_{ 1({\hat 3})} $ and  $ L^+_{ 1({\hat 3'})} $  are two
  solutions of the FRT equations related to the quantum spaces  ${\hat 3}$
  and      ${\hat 3'}$ ,respectively  then    $ L^+_{ 1({\hat 3})} \cdot
  L^+_{ 1({\hat 3'})}
 =\Delta ( L^+_{ 1({\hat 33'})}) $ is also a solution,
  which thereby defines the
 coproduct structures for the elements of  $ L^+$. Similarly those related to
 $ L^-$
 can be
  obtained. The importance of such  coproduct  is that
 it is neither related directly to the intertwinning property
  nor to the
 universal  {\cal R}-matrix and  reflects only the
 underlying Hopf algebraic structure.
 Therefore such coproducts may exist  even when the universal $ {\cal R}$ may
 not be found. We shall use this fact later for deriving
  coproduct structure
 of a new algebra (3-4), for which $ {\cal R}$-matrix is not known.
 \\ \\
 {\bf  Scheme for generation of integrable models}:

 \noindent
  $\circ$ \quad  Start with an universal $ {\cal R}$-matrix and
  reduce it to obtain
 $ R^{\pm}_{BGR}, L^{\pm}$  etc giving the objects of FRT algebra.
                                         \\  $\circ$ \quad
 Determine the
  underlying quantised algebra  and the related
 coproducts, as described above.
                             \\  $\circ$ \quad
 Yang-Baxterise
  the FRT algebra, i.e. include spectral parameters to construct
 {\it Lax operators} $L(\lambda)$ and {\it quantum } $R(\lambda,\mu)$-matrix
 (trigonometric type) of an integrable ancestor model. For our case such
 Yang-Baxterisation may be given in the form \\   $      \qquad
    R (\lambda-\mu ) ~=~ \frac {\eta}{ \xi } R^+ -
      \frac{ \xi}{\eta} R^-  ~,\ \ \ \
  L (\lambda)  ~=~ \frac {1}{ \xi  }  L^{(+)} + \xi L^{(-)} ~
 , \ \ \  \ \xi=e^{i\lambda}, \eta=e^{i\mu}.   $  \\
  \\  $\circ$ \quad
 Find different
  realisations of the   quantised algebra which  reduce the ancestor
 model to  generate different quantum integrable systems. The realisations
 in bosonic variables
    $(u,p)$ or $(
   \psi,\psi^\dagger)$ with
  $[u,p]=1$ , $
   [\psi,\psi^\dagger]=i$
 or
  $q$-bosonic variables: $(A, A^\dagger) $   with
 $~~AA^{\dagger } -  q^{-1} A^{\dagger }A ~=~ q^N~
   $ seems to
 yield physical models.
              \\  $\circ$ \quad
 Since new universal {\cal R}-matrix can be obtained from the original one
 by `twisting'
 : $   \tilde {\cal R}=    {\cal F}^{-1} \      {\cal R} \ {\cal F}^{-1}$
   , it yields in its turn  new forms of
  $ \tilde R^{\pm}_{BGR}, \tilde L^{\pm} $ giving
  new ancestor Lax operator with deformed trigonometric
  quantum $R$-matrix, which would yield  finally
 new classes of integrable models.
              \\  $\circ$ \quad
 At $ q \rightarrow 1$ , when the underlying quantum algebra reduces to
 Lie algebra the above scheme also  reduces consistently to generate
 corresponding classes of integrable models with rational quantum $R$-matrix.
 Thus in a systematic
  way we may construct directly the Lax operators along with the
 quantum $R$-matrices  of different families of
  exactly integrable quantum systems at the lattice level.

 However before moving further we should focus first on a shortcomming
 in our scheme
  by  noticing  that,  if we
 start from the explicit form of universal ${\cal R}$-matrix (1)
 and follow the above scheme literally, we will end up by
  discovering only a particular class of integrable models like sine-Gordon,
 deformeed sine-Gordon (and nonlinear Schrodinger equation at $ q=1$ limit)
 , which are related to the standard quantum group algebra (QGA)
 $ U_q(sl(2))$.
 However, we could overcome  this difficulty at the cost of loosing some
  generality in our scheme, which on the other hand
 would stimulate discovery of  a new type of
 quantised algebra  worthy of
 attention.
 Notice that at the reduced level  UYBE
  (2), acquire some extra freedom, which allows a more general choice
     of
  $ R^+_{1{\hat 3}}=
  L^+_{1({\hat 3})}$ than that obtained
    directly  from the universal ${\cal R} $ (1)
     ( which gives only the  known FRT form
  related to standard QGA, as mentioned above).
  In fact  one may choose $L^+$  in a general `upper-traiangular'
  form  as
  $\quad
 L^{(+)} = \left( \begin{array}{c}
   {\tau_1^+} \quad    {\tau_{21}} \\
     \quad \quad   {\tau_2^+}
           \end{array}   \right), \quad
  $ while   $L^-$-matrix as   lower-triangular one:  $
 \quad L^{(-)} = \left( \begin{array}{c}
  {\tau_1^-} \quad  \quad   \\
   {\tau_{12}} \quad {\tau_2^-}
           \end{array}   \right) \quad,
  $
  with yet undefined operators $\vec {\bf \tau}$. Note however that the
  forms of $R^{\pm}_{12}=R^{\pm}_{BGR} $  are
   kept unchanged, which keep track of the starting
     universal ${\cal R}$. Interestingly,
   such $L^{\pm}$  operators yield
   now from  FRT relations a new quadratic
   algebra
 \begin {eqnarray}
 \left[ \tau_{12}, \tau_{21} \right] ~=~
 - (q-q^{-1}) \left (  \tau_1^+ \tau_2^-  -  \tau_1^- \tau_2^+ \right )
  ~, \\
 \tau_i^{\pm }\tau_{ij} ~=~q^{\pm 1 } \tau_{ij} \tau_i^{\pm }~,~\ \
 \tau_i^{\pm }\tau_{ji} ~=~q^{\mp 1 } \tau_{ji} \tau_i^{\pm }~,~
 \end {eqnarray}
 %(3-4 )
 for $i,j=(1,2) ; \ q=e^{i\alpha}$,   with the Casimir operators
 \begin{equation}
 D_1= \tau_1^+ \tau_1^- \ , \  D_2= \tau_2^+ \tau_2^-  , \
 D_3= \tau_1^+ \tau_2^+ \, \
   D_4=
 2 \cos {\alpha} \left (  \tau_1^+ \tau_2^-  +  \tau_1^- \tau_2^+ \right )
 - \left[ \tau_{12}, \tau_{21} \right]_+
 .\end{equation}
 %(5)
  The related coproduct structure ($\Delta$) , antipode ($S$) and the
   counit ($\epsilon$) may be given
  respectively by
 \begin{equation}
 \Delta(  \tau_i^{\pm })=\tau_i^{\pm } \otimes \tau_i^{\pm } ,\
 \Delta(  \tau_{21})=\tau_1^{+ } \otimes \tau_{21 }
                  +   \tau_{21}) \otimes \tau^+_{2 }
 , \ \Delta(  \tau_{12})= \tau_2^{- } \otimes \tau_{12 } +
   \tau_{12}) \otimes \tau^+_{1 }
 ,\end{equation}
 %(6)
   \begin{equation}
   S (  \tau_{21})= - (\tau_1^{+ })^{-1} \tau_{21 } (\tau_2^{+ })^{-1}, \
   S (  \tau_{12})= - (\tau_2^{- })^{-1} \tau_{12 } (\tau_1^{- })^{-1}, \
 S (\tau_i^{\pm })  = (\tau_i^{\pm })^{-1},
   \end{equation}
   %(7)
   and $ \epsilon  (\tau_i^{\pm }) = 1, \epsilon (\tau_{ij}) = 0 $,
 which  endow
  this algebra with Hopf algebraic properties, though quasi-triangularity
   and the existence of  universal ${\cal R}$-matrix
    seem not   to hold for it.
  This quantised algebra may be considered as an extention of the
  trigonometric Sklyanin algebra (TSA) and  for the particular realisation
 $ \
   ~\tau_1^+ ~=~ \tau_2^- = q^{-H}, \ \
  ~\tau_1^- ~=~ \tau_2^+ =  q^{H}
 , \ ~\tau_{12} = (q-q^{-1}) X^+, \ \tau_{21} = (q-q^{-1}) X^- ~
  ,
 \ $  it reduces to
  the standard QGA.  However the extended TSA allows
   in general  a significantly richer spectrum of
     different other realisations, which
  as we see below can yield   much wider classes
  of quantum integrable  models, not reachable directly from the
  standard QGA.
  \\ \\  {\bf Explicit construction of different classes of
  quantum integrable models}:

 Not going into details of the
 explicit  calculations involved in the above scheme
 for  generating  known as well as new
 integrable quantum systems,
  we mention  only the
 main results obtained by us [2] and answer the questions raised above.
 Our results show that
 for  the first class of models, which starts from  (1),
  the generalised ancestor Lax operator
  is associated with the trigonometric $R$-matrix
  of the six-vertex model and at different realisations it yields
  field models like [2,3]\\
                       {\it 1.
  Sine-Gordon (SG) model  \\
 2. Liouville model (LM)   \\
 3. A derivative NLS (DNLS) model} [2b)]\\ {\it
 4. Massive Thirring model (MTM) }   (obtained as a fusion of DNLS [2e)]) \\
  and exactly integrable quantum
  lattice models like  \\
 {\it 1. $XXZ$-spin chain\\
 2. Lattice SG (LSG) model \\
 3. Lattice Liouville model (LLM)  \\
 4. Discrete DNLS and Distrete  MTM } in  {\it bosonic} as well as {\it
  $q$-bosonic realisations} [2e)] etc.

 At $q=1$ limit the $R$-matrix reduces to its rational form and the
  corresponding
 ancestor model yields now \\
 {\it 1.
  Nonlinear schrodinger equation (NLS) and its
  integrable lattice variant (LNLS) \\
 2. Toda chain   (TC)     \\
   3. $XXX$-spin chain} etc.

 Remarkably
 in addition to the above set of quantum integrable models another
  important class
 is generated if we  start  from a  transformed
   universal $\tilde {\cal R}$-matrix , which may be obtained  from (1)
 by twisting with $ {\cal F}= e^{\theta (Z \otimes H-H \otimes Z)}$,
 where $\theta$ is a parameter and $Z$ a central element of QGA.
        Two different situations may now occur depending on the
         representation of $Z$.

 i)  If under the
 reduction of the quantum spaces to finite dimensions, $Z \rightarrow
  {\bf I}$,
 the resultant quantum $R$-matrix becomes `$\theta$-deformed' trigonometric
 $R$-matrix  and the corresponding ancestor model yields  set of integrable
 systems like [2,3]
     \\ {\it
 1. Ablowitz-Ladik model (ALM)  \\
 2.  $6V(1)$  spin chain \\
 3. Relativistic (quantum ) Toda chain (RTC)} [2a)]  \\
 4. {\it $\theta$-deformed LSG,LLM,discrete DNLS,MTM } etc.\\ and
 also at the $q=1$ limit :  {\it $\theta$-deformed LNLS,Toda chain} etc.

  ii) However if at finite
 dimensional reductions of $ {\hat 1 }$ and ${ \hat 2}$
  the central element $Z $ has
  eigenvalues  $\lambda$ and $\mu$ , respectively
  , it induces a `colour' BGR as finite dimensional representation
  of the twisted universal {\cal R}-matrix.
 This necessitates the formulation of `colour' FRT algebra etc. resulting
 finally integrable models with a new nonadditive spectral
  parameter dependent quantum $R$-matrix [2 c)].
 \\ \\
 $\diamond$  Thus  following our scheme we are able to
  construct integrable  ancestor
  Lax operators
     and through
 consistent realisations  generate various families
 of quantum integrable models representing  known as well as new lattice
  and field models.
            \\   $\diamond$
  The models sharing the same quantum
   $R$-matrix are found to be the descendants
 of the same ancestor model, which  thus explains  nicely their `commonness'
 and answers the question raised here.
 \\ $\diamond$
   The criterion for integrable nonlinearity seems to be dictated
 by different concrete realisations of the
  quantised algebra (3-4) through physical variables.
  Transition to  continuum
 limit distorts such `exact'
      nonlinearity.

   Extention of the scheme presentated here
   to higher dimensional ( $(N\times N)$) Lax operators  and also to
  other algebras like rotational,
  projective as well as
 supersymmetric algebras
 should be important  directions of development.

 Support of   Alexander von Humboldt Foundation fellowship
  is sincerely acknowledged
 by the author.
  \\ \\  {\bf
 References}:\\
                \\
 \ [1] \ N.Yu.
 Reshitikhin , L.A. Takhtajan  and L.D. Faddeev  1989 {\it Algebra
 and analysis }\\ \ {\bf 1} 17 \\ \\
 \ [2] \ A. Kundu  and B. Basumallick \  a) \
  1992
  {\it Mod. Phys. Lett.  }
  {\bf A 7 }  61 ;
  \\ \ $ \qquad $ b)                   \
   1993 {\it J. Math. Phys.}  {\bf 34} 1052 ;
 \\ \ $ \qquad $ c) \
    1993 {\it
 Coloured FRT algebra and its Yang-Baxterisation
 leading to integrable models with nonadditive $R$-matrix}
  Saha Inst. preprint , April 1993;\\ $\qquad $
 B. Basumallick and A. Kundu  \ d) \ 1992 {\it J. Phys.  }  {\bf A 25 }  4147
 ; \\ \ $ \qquad $ e) \ 1992 {\it  Phys.Lett.  }  {\bf B 287 } 149
 \\                                \\ \
 [3] \ P.P. Kulish and E.K. Sklyanin  1982,
 in {\it  Lecture notes in Physics }
  (Springer Verlag, Berlin) {\bf  151}  61.

 \end{document}